# "COMPUTER SAYS NO": ALGORITHMIC DECISION SUPPORT AND ORGANISATIONAL RESPONSIBILITY

ANGELIKA ADENSAMER, RITA GSENGER, AND LUKAS DANIEL KLAUSNER

ABSTRACT. Algorithmic decision support is increasingly used in a whole array of different contexts and structures in various areas of society, influencing many people's lives. Its use raises questions, among others, about accountability, transparency and responsibility. While there is substantial research on the issue of algorithmic systems and responsibility in general, there is little to no prior research on *organisational* responsibility and its attribution. Our article aims to fill that gap; we give a brief overview of the central issues connected to ADS, responsibility and decision-making in organisational contexts and identify open questions and research gaps. Furthermore, we describe a set of guidelines and a complementary digital tool to assist practitioners in mapping responsibility when introducing ADS within their organisational context.

## 1. Introduction

Algorithms are changing work environments more and more every year. While discrimination and transparency are – quite rightly – discussed a lot in academic literature, questions of responsibility in organisations are rarely addressed in sufficient detail. In this paper, we want to offer tools for organisations to deal with and preempt common problems of responsibility that occur when introducing and using algorithmic decision support (ADS; sometimes also algorithmic decision-making, ADM) in organisational contexts.

ADS is used in more and more domains every year (e. g. human resources (Dreyer and Schulz 2019, p. 7), access to welfare and credit (O'Neil 2016), policing (Bennett Moses and Chan 2018; Ferguson 2017) and sentencing (Christin 2017)), affecting decisions with profound influence on people's lives. This kind of technology is often perceived as particularly capable of making more objective and value-neutral decisions by applying more processing power and using more data than any human being ever could, which is stipulated to be an advantage for decision-making (Christin 2017; Gillespie 2016). Moreover, algorithmic systems are thought to be able to "'cure' experts from their own subjective weaknesses" (Christin 2016, p. 31). While using ADS systems can have benefits, such as in the field of medical imaging or diagnosis (Castellucia and Métayer 2019), the entire process involves many different kinds of potential and actual unequal treatment, from biases in the underlying data (Barocas and Selbst 2016; Hao 2019) through implicit (Friedman and Nissenbaum 1996) and explicit assumptions made when designing the model (Bennett Moses and Chan 2018, p. 809 ff.) to how the predictions and "recommendations" are explained and presented to the end user (Goddard, Roudsari, and Wyatt 2011).

Moreover, the use of this technology influences human operators and the decisions they make when being guided by an ADS system (Lee 2018; Prahl and Van Swol 2017). Promises of higher efficiency and "better" decisions stand in contrast with problems of transparency, explainability and, fundamentally, the question of responsibility. While in many work environments the people tasked with making decisions supported by algorithms retain the final responsibility therefor, in most cases they do not have a choice or say in whether and how to employ ADS systems in their work. This raises questions and conflicts of interests between different levels of organisational hierarchies, with possible impact on the livelihood of people without the agency to exert relevant influence on these decisions.

Much of the previous research on this topic has focussed on the consequences of algorithmic or algorithm-supported decision-making for the people impacted by these decisions, while its effects on the working conditions of decision makers have received less attention. Many employees in middle management and civil servants with executive or decision-making powers (in short: decision makers) are about to find themselves (or already are) in positions where they are still responsible for the consequences of their decisions after the introduction of an ADS system. At the same time, they are expected to not only make decisions more efficiently (Vieth and Wagner 2017, p. 20), faster and with a higher throughput (Zweig, Fischer, and Lischka 2018, p. 15) than before the introduction of ADS, but also to test and validate decisions "suggested" by an algorithm they might only partially understand at best. In contrast to human decision-making, which almost always allows at least some leeway for discretionary decisions and judgement calls, algorithms by necessity do not make this allowance (Zweig, Fischer, and Lischka 2018, p. 15).

Increased use of algorithmic support in decision-making also exacerbates the risks of discrimination and inequality, both on an individual level (since the expected higher throughput rates afford the decision makers less time to consider any possible discriminatory impact their decisions might have) as well as on an institutional level, as biases and unfairness in ADS systems have the potential to affect a substantially larger number of people than any individual decision maker's biases. In our previous work focussing on algorithmic systems in policing and justice (Adensamer and Klausner 2021a), we coined the distinction between "retail bias" and "wholesale bias". The former refers to the bias and discriminatory effect caused by individual human decisions; their impact – while possibly substantial – nonetheless remains limited in scope, timeframe, or location. The latter describes subjecting larger groups of people to the same automated decision mechanism (and its according biases), with potentially categorically more extensive repercussions, affecting orders of magnitudes more cases and people with far fewer constraints and limitations. Moreover, the ostensible objectivity and neutrality can facilitate discriminatory results being denied or explained away, which in turn can lead to reinforced discrimination and can (together with lacking or insufficient transparency of the algorithms involved) make avoiding or reducing unequal treatment even harder. All of this presents additional and novel barriers to equality, inclusion and freedom from discrimination.

In this paper, we focus on the aforementioned and hitherto mostly neglected topic of ADS and working conditions and in particular the question of organisational responsibility for the introduction and use of algorithmic decision support.

The concepts of "accountability" and "responsibility" are often defined referencing each other and are by no means clearly delimited in academic literature (McGrath and Whitty 2018). The terminological debate is ongoing; for instance, Koppell (2005, p. 96 ff.) considers responsibility (in the sense of "being guided by rules, standards and norms and observing them") to be one of five dimensions of accountability (together with controllability, liability, responsiveness and transparency). Furthermore, Koppell rejects a unitary definition of the concept.

Accountability commonly refers to individuals needing to "explain (account) to those who are entitled to ask (e. g. regulators or a patient) for their decision-making process" (H. Smith 2021, p. 5); accountability thus is connected to being called to account, in contrast to being (abstractly) responsible for an action (Cornock 2011). Following this understanding, accountability includes concepts such as "answerability, blame, burden and obligation" (Ieraci 2007, p. 63), whereas responsibility is related to "trust, capability, judgement and choice" (Ieraci 2007, p. 63). Therefore, responsibility is more strongly related to doing and might be taken on by individuals themselves, whereas accountability involves reporting and is bestowed upon another person (Ieraci 2007).

To define and delimit the concepts of accountability and responsibility, McGrath and Whitty (2018) conducted a literature review and an analysis of the terms' usage, resulting in a definition of responsibility as "an obligation to satisfactorily perform a task" (p. 701) and accountability as "liability for ensuring a task is satisfactorily done" (p. 702). Accountability arising from legislative or organisational sources cannot be delegated in most circumstances; responsibility, however, can be. In an organisation, staff are responsible for certain tasks and duties, for which managers will hold them accountable. In that context, accountability can only be identified relative to a certain task and transitions between levels of an organisation. Overall, accountability does not necessarily originate from legal, organisational or contractual sources alone, but also from informal ones; social groups, for instance, can hold their members accountable with measures such as correction or exclusions (McGrath and Whitty 2018). In the context of algorithmic systems, a fundamental precondition for a person to be attributed responsibility is that they need to have the skills and knowledge required to comprehend and critically question a system's decision-making.

Following these distinctions, in this paper we distinguish between "organisational responsibility" and "organisational accountability" by considering organisational accountability to describe liability, blame and burden within the entire sociotechnological system surrounding organisational ADS use (including, but not limited to the legal framework), whereas organisational responsibility is concerned with questions of attribution, choice and obligation, focussing especially on the organisational embedding of ADS systems.

We analyse the different levels of an organisation and the associated degrees of agency and responsibility of human decision makers related to the processes of introduction, usage and evaluation of ADS systems. We then describe a method for mapping degrees and fields of responsibility and accountability in these processes to the various actors and roles in such an organisation, thus providing a model that can point to discrepancies and problems as well as offer guidance on how to avoid them. As a practical contribution towards solving the problems we identify, we finally present a set of guidelines for employers and works councils (Adensamer and Klausner 2021b) highlighting possible areas of concern regarding questions of responsibility when introducing ADS systems, augmented by a supplementary prototyped digital tool called "VerA"[1] to assist with responsibility mapping and highlighting potential conflicts of interest and responsibility vacuums. Both the guidelines and VerA were developed in cooperation with the Vienna Chamber of Labour and in close discussion with experts from academia and industry to ensure their practical relevance and utility.

## 2. Organisations, Accountability and Responsibility

Algorithmic accountability has become an area of growing research interest in recent years. Current research draws upon both the existing work from the field of accountability theory (in organisation theory) as well as the study of fairness, accountability and transparency in computer science. Wieringa (2020) has recently given a systematic and structured overview of the current state of research. She points to and follows Bovens' widely used definition and ontology of accountability (Bovens 2007), which considers "accountability" to be a social relation composed of (in Wieringa's exposition) five constituents: (1) an actor, (2) an accountability forum, (3) the relationship between actor and forum, (4) the content, parameters and framework of the account and finally (5) the kind of consequences which may result for the actor. Her analysis finds that algorithmic accountability is best described as "a networked account for a socio-technical algorithmic system, following the various stages of the system's lifecycle" in which "multiple actors [...] have the obligation to explain and justify" their actions outside and within the system in question (Wieringa 2020, p. 10). In particular, our approach aligns strongly with her finding that responsibility is distributed between these different actors and requires both accounts within the organisation and accounts to external fora. (We will return to this issue and build upon Wieringa's work in section 5.)

In the field of organisation theory itself, however, the question of organisational responsibility for the use of ADS (or algorithms, more generally) has been less in the spotlight. Research in organisation theory has focussed mainly[2] on the change of organisational structures through the use of the internet, in particular the effect of evaluations (Curchod et al. 2020), and the corresponding change in role relations (Barley 2015) or occupation boundary dynamics (Barrett et al. 2012). Notable exceptions to this gap are the recent works of Faraj, Pachidi, and Sayegh (2018), Kellogg, Valentine, and Christin (2020) and Moradi and Levy (2020). Faraj, Pachidi, and Sayegh (2018) investigate the effect of the introduction of learning algorithms on organisations, work and organising. They predict, on the one hand, a continuation of the existing trends towards automation, productivity gains, job gains and losses and the emergence of new types of occupations, but on the other hand contend that a combination of novel factors (black-boxed performance, comprehensive digitisation, anticipatory quantification and hidden politics) will likely raise additional and unanticipated challenges, such as the transformation of the concept of professional expertise, fundamental conversions of existing occupations and their boundaries, and new and increased opportunities for both control and coordination. Kellogg, Valentine, and Christin (2020) discuss a structured ontology of mechanisms which the introduction of algorithmic systems in the workplace makes available as tools for employers to algorithmically control their workers: Per their analysis, algorithms allow employers to direct workers by restricting and recommending, evaluate them by recording and rating and discipline them by replacing and rewarding. Based on this analysis, they then discuss four key issue clusters showing ways in which algorithmic control is a novel field of rational control (the problematic focus on the economic value of algorithms; how algorithmic systems afford employers a new form of control over their workers distinct from previous methods; the emerging landscape of occupations fundamentally defined by their dependence on or interaction with algorithmic systems; and individual and collective forms of resistance to algorithmic control) and explore possible future research directions. Finally, Moradi and Levy (2020) first discuss ways in which many forecasts of the effect of AI and algorithms on the future of work restrict themselves to viewing work through a purely technical lens, neglecting social, organisational and contextual aspects of work. They then give their own, more holistic prediction of the changes the widespread introduction of algorithmic systems into the workplace will cause, which they argue can be summarised as shifting risks from companies and employers to workers (through increasing the flexibility of staffing and scheduling, redefining compensable work, detecting and preventing loss and fraud, and incentivising and exhausting productivity). They argue that these different methods have in common that existing inefficiencies within an organisational structure are not removed, but that ADS instead only serves to "redistribute the risks and costs of these inefficiencies to workers" (Moradi and Levy 2020, p. 278).

---

[1] https://vera.arbeiterkammer.at (accessed 15 June 2021)
[2] But not exclusively – note, for example, the recent critical reflection by Heimstädt and Dobusch (2020) on the fundamental limitations in organisation theory's existing approaches for investigating the interplay between accountability and transparency.

More broadly speaking, research and public discourse on the effects of ADS on human decision-making, working conditions and organisation thus far have generally paid relatively little attention to the question of organisational responsibility in our sense of the term. To give just a few examples: While a study published by the Council of Europe on discrimination, artificial intelligence and ADS contained several important recommendations, proposed safeguards and regulations to ensure fairness and protect human rights, none of them were geared towards the role of decision makers using ADS (Zuiderveen Borgesius 2018). The Bertelsmann Stiftung working paper "Wo Maschinen irren können" ("Where Machines Can Err") makes several important observations on what types of errors ADS systems can cause, with mistakes arising from emergent phenomena (such as behavioural adaptation by the decision makers as they become used to utilising the ADS system) considered particularly difficult to prevent or mitigate (Zweig, Fischer, and Lischka 2018, p. 27). The authors identify the role of decision makers and their working conditions as crucial to the topic and make some recommendations, such as supplementing the ADS system with "instruction leaflets" explaining the underlying data and modelling assumptions and intended use cases as well as listing the algorithm's known "side effects". Nonetheless, while their analysis outlines a chain of consecutive areas of responsibility in the use of ADS (based on their five-phase conception of ADS systems), it neglects the essential question of the fundamental shift in responsibility effected by the introduction of ADS systems in the first place (and also pays relatively little attention to demarcation problems in assigning responsibility at the phases' margins).

Finally, while studies, policy papers and white papers on risks and benefits and the ethical questions of using ADS (such as Christen et al. 2020; Fjeld et al. 2020) with a wide scope often explore the issues of accountability, liability and responsibility, they mostly neglect to specifically investigate the intricacies of divided competences within complex and often highly hierarchical organisations and the questions this raises regarding (intra-)organisational responsibility. This issue is related, but not identical, to the so-called "many-hands problem" (see e.g. Poel, Royakkers, and Zwart 2015; Thompson 2017), in which responsibility can be attributed to a collective of actors, but distributing it among them or ascertaining individual culpabilities remains impossible. In our framing of organisational responsibility, we argue that the corresponding problem is the existence of a "responsibility vacuum"; see the following section for more on this.

## 3. Decision-Making and Algorithmic Decision "Support"

We now want to turn our attention to the practical effects and consequences introducing ADS systems has on decision-making, and in particular the ramifications for the employees using them. The introduction of ADS risks conflicts of interest between different levels of organisational hierarchies and in particular might affect the work lives of those employees who lack the possibility to exert influence on the decisions surrounding the use of automation. One key issue is that the decision on whether to introduce automation is not taken by the employees who consequently have to use the technology in their work (cf. Faraj, Pachidi, and Sayegh 2018, p. 366 f.). In a study by Christin (2017) on the introduction of ADS systems in the criminal justice system and in web journalism, managers generally emphasised the effectiveness of the introduced systems and the up-to-dateness of their enterprise. However, employees in these studies tried to avoid using the ADS systems as much as possible, relying on their own decision-making capabilities instead. Irrespective of whether employees have any say on the introduction of automated systems such as ADS, such an event can change the expectations placed on their work, in some cases fundamentally so: At the very least, they are expected to make decisions with higher efficiency (Vieth and Wagner 2017, p. 20) and more quickly (Zweig, Fischer, and Lischka 2018, p. 15), while still being held responsible therefor. Decision makers often have to assess and then accept or decline automated "suggestions" given to them by an algorithmic system (irrespective of whether they have been explicitly trained to understand the model or whether they are given the necessary documentation). In addition, the use of ADS can markedly increase the potential of discriminatory decisions (through what we have named "wholesale bias", see above in section 1 and Adensamer and Klausner (2021a, p. 4)).[3]

Recommendations by ADS systems influence human decision-making and can introduce various different effects and behaviours such as complacency, biases and aversion. Especially in work contexts where humans were previously responsible for decision-making on their own, but where decisions are now made using ADS systems, decision makers often do their best to circumvent the systems, if possible (Christin 2017), or accept the decisions without questioning them if they have little say (Schäufele 2017). Their perceptions of and attitudes towards ADS systems have been found to have significant impact on the behaviours in interacting with them. Users' attitudes additionally vary according to cultural influences (Hoff and Bashir 2015), their knowledge about the systems (Alexander, Blinder, and Zak 2019; Burton, Stein, and Jensen 2020) and the opinions exhibited by peers and the media (Burton, Stein, and Jensen 2020). We will discuss a few important effects in the use of ADS systems in the sequel.

---

[3] See also Loi and Spielkamp (2021) for a review of regulatory guidelines on AI in the public sector, whose introductory analysis on the challenges for accountability independently comes to similar conclusions regarding the most prominent phenomena as our article.

First, human operators or users might be too complacent regarding the suggested decisions of an ADS system, leading to errors in judgement. Complacency occurs when human operators do not monitor a system sufficiently closely, e.g. due to an explicit or implicit assumption of a satisfying system state (Parasuraman and Manzey 2010). Moreover, complacency is particularly likely when operators only passively monitor an automated system and only take action if something goes wrong, which risks negligence, delayed reactions and misuse (Bahner, Hüper, and Manzey 2008; Zerilli et al. 2019). Complacency – in contrast to other biases, such as automation bias (see below) – also occurs if systems do *not* give advice even when they should do so (Wickens et al. 2015). Complacency in the context of ADS systems has been researched by Bahner, Hüper, and Manzey (2008) in a laboratory study in which participants were tasked with identifying faulty recommendations made by an ADS system providing advice on fault diagnosis and management, after the system had previously worked well. All participants proved to be complacent towards the system to some extent, which was explained to be caused by inefficient information sampling behaviour. Moreover, commission errors occurred in cases of high levels of complacency. Automation complacency in the form of failure to question the system has been observed in practice e.g. in ADS systems used in policing (Harcourt 2007) and social work (Eubanks 2018).

Second, automation bias with regard to the reception of recommendations from ADS systems has been observed. Similar to complacency, automation bias occurs when a human is interacting with a system which is giving incorrect advice, but the human decision maker does not question the suggested decision or seek out additional information (Parasuraman and Manzey 2010). Humans have bounded rationality, i.e. a limited capacity to process information (Lack and Rousseau 2016), and are, according to social psychology, cognitive misers (Clarke 2007), which means they prefer simpler solutions and explanations irrespective of their intelligence. Humans thus have a tendency to try to save cognitive effort and mental resources when making decisions, especially by reducing time and by avoiding cognitively strenuous activities, such as task switching (Dunn and Risko 2019). These types of mental shortcuts or heuristics lead to individuals allocating responsibility about decision outcomes to automated systems more easily, resulting in automation bias (Parasuraman and Manzey 2010). Various cognitive biases can lead to erroneous judgements or disadvantageous choices (Kahneman 2003; Kahneman and Klein 2009; Tversky and Kahneman 1974) due to the use of heuristics in decision-making (Mosier and Skitka 1996). Automation bias refers to a type of bias specifically resulting from interaction with automation (Parasuraman and Manzey 2010).

Moreover, the notion of increased working capacity, memory and objectivity of ADS systems influences the tendency to allocate responsibility to the system. When working in a group, individuals have a tendency to give responsibility to others, which is also applicable to working with an automated system. Feier et al.'s recent empirical laboratory study (Feier, Gogoll, and Uhl 2021) implies that delegating tasks to algorithms and machines instead of to other people allows decision makers to more easily shift the blame for negative outcomes. Automation bias might lead to using "follow incorrect recommendations by an ADS system" as a heuristic replacement for seeking information (Parasuraman and Manzey 2010); the bias can be reduced by decreasing the cognitive load of human operators (Lyell and Coiera 2017). Furthermore, individuals who feel they are held accountable for their decisions examine them more thoroughly and include more information in the decision-making process, thus reducing the influence of automation bias. Consequently, ensuring that users of ADS systems have direct contact with the people affected by the system's suggestions might entice decision makers to examine the decision suggestions more thoroughly and thus increase overall accountability (Skitka, Mosier, and Burdick 2000).

Finally, one more type of bias concerns the distinction between algorithmic and human recommendations and suggestions. Individuals often prefer human recommendations over those of ADS systems and estimate a human's input to be more valuable. Furthermore, professionals who use ADS are frequently more harshly judged than human beings, especially if they make mistakes. The phenomenon of people not using algorithmic systems even if doing so would be beneficial is referred to as algorithmic aversion (Dietvorst, Simmons, and Massey 2015; Dietvorst, Simmons, and Massey 2018). The degree of aversion against an automated system depends on the type of decision and task in question (Castelo, Bros, and Lehmann 2019). In a study by Lee (2018), an algorithmic system was used to decide about shift scheduling, i.e. used for managerial tasks. Two types of decisions were contrasted in the study: decisions involving "human" skills and others requiring "mechanical" skills. The former included scenarios involving work assignment and work scheduling, the latter scenarios involving hiring and work evaluation. Participants judged the algorithm's and human managers' decisions about the mechanical tasks as equally fair and assumed that the algorithm was unbiased and following a set of rules, which each worker should equally be subjected to. However, some participants pointed out that the algorithm was not able to understand nuances of individual workers' life situations, possibly decreasing its capability to make good decisions. Humans' decisions were estimated to be more fair than the algorithm's decisions regarding hiring workers and evaluating their performance, since the human managers would be – according to the study participants – capable of identifying the best candidates according to their merit, even though it was sometimes judged to be subjective. However, participants did criticise the lack of intuition and judging those human qualities which are hard to quantify by an ADS system. Moreover, evaluating work performance was regarded by participants to be fairer if done by a human manager, due to skill and authority ascribed to them. Finally, the algorithm's judgement about work performance was deemed unfair, as the algorithm cannot take contextual information – such as a worker having a "bad day" – into consideration (Lee 2018).

Furthermore, expectations in the capabilities of the ADS system play a role regarding the perception and use of such systems (Burton, Stein, and Jensen 2020) – and ultimately, their success. Aversion against ADS systems is especially high if participants do not know how the system works, which changes when participants learn more about the system's functioning (Yeomans et al. 2019). Aversion can furthermore be reduced if individuals can exert at least some influence on the system's decisions, even if their choices are very limited, as found by Dietvorst, Simmons, and Massey (2018). Overall, it has been found that trust in automated systems deteriorates more rapidly than trust in human beings if mistakes occur (i.e. mistakes by machines are penalised more strongly). Individuals are more confident in humans' ability to adapt their choices and correct their errors (Prahl and Van Swol 2017). However, some studies conversely show a clear preference for recommendations made by ADS systems, especially before mistakes are made known to participants (Dietvorst, Simmons, and Massey 2015; Logg, Minson, and Moore 2017; Prahl and Van Swol 2017).

Even as human beings are prone to biased decisions and heuristics, they are often at least partially aware of these processes and by no means "happy fools who blindly answer erroneous questions without realising it" (De Neys, Rossi, and Houdé 2013, p. 269). Some operators using ADS systems do act appropriately, question unexpected suggestions and decisions and bring their own judgement into play. In a study by De-Arteaga, Fogliato, and Chouldechova (2020), the influence of an accidentally faulty ADS system used to help call workers decide on screening decisions for possible child maltreatment cases was investigated. They found that workers were influenced by the recommendations by the ADS system, but did not succumb to either automation bias or algorithmic aversion; according to the study, they made reasonable choices about which (erroneous) algorithmic classifications to overrule and which to accept. Conversely, however, Kolkman (forthcoming) found through ethnographic work on the use of several algorithmic models in practice that transparency of algorithms is "at best problematic and at worst unattainable" even for experts, i.e. people working with models professionally.

Notably, despite the fact that one of the purported intended effects of introducing ADS is to improve decision-making, Skitka et al. find that "remarkably few" studies have been conducted to examine the question of whether the introduction of ADS does actually reduce the error rate, and that on the contrary, available evidence at least allows for the possibility that the actual effect is not an unmitigated reduction in human errors, but instead a partial replacement of human-type errors with new (algorithmically induced) types of errors (Skitka, Mosier, and Burdick 1999, p. 992).

## 4. Shifts and Changes in Responsibility

The introduction of ADS systems in work environments can have significant impacts on the distribution of responsibility, agency and liability. Often, power (and sometimes also responsibility) is shifted away from employees (who previously made their decisions without using ADS while having more time to devote to each individual case) and to external developers who know less about the day-to-day application and the impact of the decisions (see also Moradi and Levy (2020, p. 278) for more on risk-shifting and Wagner (2019) on liability and criteria for meaningful agency in quasi-automated systems). These shifts are likely to lead to problems if not addressed properly.

First, it can cause problems when new tasks and responsibilities are shifted to employees who are not trained for them. The job description of case workers can change drastically when an ADS system is introduced, e.g. from evaluating individual case files to controlling and questioning the workings of a (possibly complicated) computer system. If the employees do not have enough knowledge and effective means to question the decisions of the ADS system, they cannot take responsibility for its outcomes. When introducing ADS in an organisation, it is thus also imperative to carefully consider the work environment of decision makers who are now "supported" by the algorithm. While their job description is likely to change (at least tacitly and by implication), their competences, knowledge and training do not automatically develop at the same time or pace. If they are not given adequate support, they might be put in an unsustainable situation, as the practical influence they have on decisions they "make" wanes while their degree of responsibility for such decisions persists.

Second, if the responsible employee is not authorised to counteract or make the necessary changes in the ADS system, they cannot be allocated ultimate responsibility for the resulting decisions. In principle, employees can be held accountable by their employers when they (illegally) discriminate in their decisions. However, when it is an algorithmic system which discriminates and the employee (directed to make use of ADS in their work) is not enabled by their employer to sufficiently understand and scrutinise the model, the underlying data, etc., the situation is rather different. We contend that employees cannot be held accountable in this case, as they do not have significant say in the functioning of the algorithmic system. Responsibility can never exceed the scope for decision-making.

Third, the assignment of responsibilities might be neglected, and what we call a "responsibility vacuum" might arise. Between the management or governmental agency deciding on the introduction of ADS, the people responsible for integrating ADS systems in the organisation's workflows, the developers creating or modifying the system, the people performing quality assurance on both the software and the underlying database, independent evaluators and auditors, systems security specialists and the actual decision makers, it often remains unclear who is actually and practically responsible for the partially algorithmic, partially human decision in a specific case. Even worse, this can split the responsibility between many different parties not only within, but possibly even outside the organisation in question. Especially for those affected by the decision, not being able to identify who is responsible for deciding their case can be frustrating or even dangerous and create situations that are difficult to navigate.

In order to prevent these problems, questions of responsibility have to be addressed before the introduction of an ADS system. We have developed a mapping tool (see below) which can help map responsibilities and identify potential problems. With or without this tool, it remains necessary to follow a structured approach wherein all actors, tasks, responsibilities and ways of communication are considered and addressed in order to eliminate oversights and gaps.

## 5. Guidelines and Responsibility Mapping Using VerA

The changes and shifts in responsibility mentioned above often happen without the necessary accompanying forethought or mitigating efforts. In our analysis, a first step to address these effects is to facilitate knowledge and awareness therefor. There are several guidelines, reports and manuals that are useful resources for these processes (such as Engelmann and Puntschuh 2020; O'Neil and Gunn 2020; Puntschuh and Fetic 2020a; Puntschuh and Fetic 2020b; Reisman et al. 2018) and give an excellent overview on how multifaceted potential problems with using ADS systems in the workplace are, ranging from costs and meeting objectives through transparency, security, data protection, discrimination and documentation to practices of assessment and evaluation. However, questions of responsibility are only touched upon superficially in all these documents, thus leaving a gap in the supporting literature – which is why we wanted to provide more detailed guidance on this topic.

As a first step towards filling this gap, we have developed (in cooperation with the Vienna Chamber of Labour and in close coordination with experts from academia and industry) a set of guidelines with a particular focus on the question of organisational responsibility. We collected the different problems related to organisational responsibility from existing research literature and practical examples and grouped them by their characteristics; in parallel, we also identified, grouped and differentiated sets of both tasks, duties and roles as well as responsibilities. We then framed the problems in terms of relationships between the tasks/duties/roles and responsibilities; this kind of structure allowed us to conduct a structured analysis of tasks/duties/roles and responsibilities, what they each should and should not comprise, and their relationships with each other. Plans to test our guidelines and the supplementary digital tool in case studies unfortunately had to be postponed/shelved due to the COVID-19 pandemic and its repercussions. In lieu of that, we discussed our analysis, guidelines and VerA with experts from several scientific disciplines (business studies, cognitive science, critical algorithm studies, human–computer interaction, organisation theory and sociology) as well as with representatives of both sides of Austria's *Sozialpartnerschaft*[4] (namely from the Chamber of Labour and the Federation of Austrian Industries). We were primarily forced and/or motivated to seek input from *Sozialpartnerschaft* representatives (instead of more direct input from specific companies with particularly relevant use cases) by COVID-19 pandemic concerns, but also hoped (fortunately, correctly) that the industrial relations representatives would have a reasonably comprehensive overview of current developments, concerns and use cases across different industries and sectors. The particular focus of our informal exchanges lay on the practicability of our guidelines and the usability of VerA.

---

[4] "*Sozialpartnerschaft*" is the term for the Austrian institutionalised cooperation governing industrial relations, i.e. the formal cooperation and exchange between the representations of both employers and the labour movement.

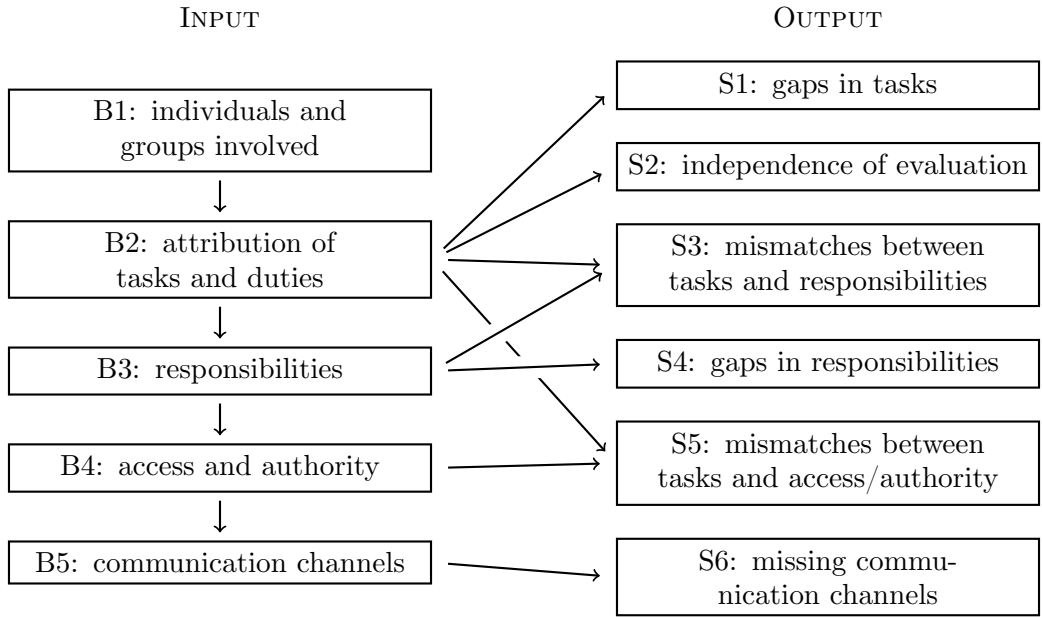

Figure 1. VerA flowchart diagram and relations between input and output.

In our guidelines we outline typical problems of shifts in responsibility (see above in section 4) and give advice on how to conceptualise and organise the different kinds of responsibility and their distribution before and after the introduction of an ADS system. It is vital to have a comprehensive overview over tasks and duties to be able to ascertain that they are all assigned and can all be fulfilled. We then offer a typology of tasks and duties to support such a responsibility mapping, broken down as follows: the fundamental decision for or against introducing an ADS system, implementation of the system into the organisation's existing processes and structure, development of the system, the practical use thereof, system security, data management and finally the (independent) evaluation of the system. Our guidelines also promote proactive handling of mistakes as well as formalising complaint and feedback channels (both internal and external) to enable the fastest possible detection of errors and to ensure timely corrections thereof. The guidelines also give a brief overview over relevant aspects from data protection law, especially concerning liability and individual rights as well as specific law regarding automated decisions on people. Moreover, we present some technical requirements necessary for adequate personal responsibility to be possible, such as understandability, transparency, documentation and modifiability of the system. Finally, a list of guiding questions is intended to help put the preceding recommendations into practice.

In addition to the guidelines, we have prototyped a supplementary digital tool called "VerA". VerA is intended to help with mapping responsibility and duties in connection with ADS systems. Naturally, the automated output produced by such a digital tool can only identify a selection of potential problems and invite the user(s) to further examine these and other possible issues; it is not a final assessment, but supposed to serve as a stepping stone towards deeper analysis and introspection.

VerA consists of five blocks in which the user answers questions on the ADS system (to be) introduced in their organisation, followed by six sections of (automated) output on potential problems of responsibility (cf. Figure 1 and Figure 2). The input queries are structured as follows:

- **Block 1** asks for the names of all individuals and/or groups who are involved with the ADS system; together with the (default) option "nobody", these are the possible answers for all blocks that follow.
- In **block 2**, users are asked to attribute tasks and duties to individuals/groups for several areas, namely: the fundamental decision for or against the introduction of the ADS system, implementation, development, actual use in practice, system security, data management and evaluation.
- **Block 3** contains questions of responsibility for several issues, i.e. who specifically has to solve problems related to and has to bear the consequences of (1) the system not meeting its set targets, (2) the system not being implemented properly into the organisation's processes and structure, (3) data protection complaints, (4) system security and integrity breaches, and (5) incorrect use of the ADS system.
- **Block 4** concerns questions of access and authority, namely: (1) Who is authorised to halt the use of the ADS system entirely? (2) Who can change the implementation and practical use of the system? (3) Who can correct and change entries in the database? (4) Who can institute security measures?
- The final **block 5** asks for the existence of communication channels and internal complaint mechanisms between the different people and groups involved.

The output consists of six sections, each of which focuses on a different set of potential problems:

- **Section 1** highlights possible gaps in tasks that are left unaddressed (by giving the answer "nobody" in block 2).
- **Section 2** addresses problems regarding the independence of the evaluation in case the people tasked with the evaluation have other tasks and hence potential conflicts of interest.
- **Section 3** points out when a person or group is responsible for an area (block 3) but is not tasked with it (block 2).
- Similar to section 1, **section 4** flags gaps in responsibility (i.e. answering "nobody" in block 3).
- In **section 5**, VerA highlights potential problems with lacking congruence between tasks/duties and access/authority: It flags areas where a person or group is responsible (block 3) but is not authorised to make changes in the corresponding area (block 4).
- Finally, **section 6** concerns problems with missing, but necessary channels of communication and complaints between different groups.

Both our guidelines and VerA are intentionally designed to be applicable to a wide range of organisations. At the same time, this lack of focus can also limit their benefit for more specific use cases. We have modelled our solutions on organisations which have some hierarchical structures (e.g. a CEO, department heads, etc. with power to decide over their subordinates and their working conditions), a non-negligible number of employees involved and at least some division of labour (i. e. assuming that the ADS system is not developed by the same people who thereafter use it in their daily work, or that different groups or people might be responsible for system security and data management); they may therefore not be applicable to organisations below a certain size, those with very flat hierarchies or nonstandard decision structures (e. g. sociocratic). However, we feel it it is still usable at least as an inspiration and motivation for introspection on questions of organisational responsibility for the use of ADS even in such use cases – as long as responsibilities, duties and authorisations are defined, at least to a certain degree (and we would consider it a potential red flag if they are not).

Figure 2. Screenshots of parts of VerA's input and output.

## 6. Limitations and Further Research

This paper provides an overview of challenges typically encountered by using ADS with biases and faulty decisions. However, some issues are outside the scope of our investigation, such as the role of trust people have in organisations and the algorithmic systems or the influence of individuals' expertise and prior knowledge. A study by the Pew Research Center in 2018 (A. Smith 2018) has shown that people are sceptical concerning the use of their data in ADS systems and have little trust in proper use by the companies developing and providing such systems. The role of operators' and users' attitudes is crucial, as is the influence of the knowledge people have about ADS systems, which can alter their attitude (see e. g. Alexander, Blinder, and Zak 2019; Burton, Stein, and Jensen 2020; Lee and Baykal 2017). Trust by employees of an organisation using ADS systems is important, however, to their responsible use thereof.

Some issues and challenges with ADS systems could not be considered in detail, but should be investigated further in future research. A considerable amount of research has pointed out that the data used for the models underlying ADS are not objective and often merely reproduce the status quo (Barocas and Selbst 2016; Christin 2017; Eubanks 2018; O'Neil 2016). Often, specific data sets are used due to ease of access and availability even if they are known to be biased or opaque. These issues can have severe consequences for an ADS system's recommendations. Furthermore, long-term predictions (e.g. on human life trajectories) have been shown to be inaccurate no matter the approach and methods used (Salganik et al. 2020), putting into question the general applicability of predictive technology beyond a very limited scope. Furthermore, algorithms have also been researched from the perspective of performativity, i.e. focussing on their influence on and interplay with social reality instead of seeing them as mostly static entities (Glaser, Pollock, and D'Adderio 2021), a perspective we could not include in this paper.

We qualified ADS systems as socio-technical algorithmic phenomena of responsibility, accountability and agency involving agents and technological artefacts. This approach has been described at length in actor–network theory (Blok, Farías, and Roberts 2020) and materialist accounts of technology (Lievrouw 2014; Sterne 2014). Materialist accounts argue for a consideration of the co-production between human agents and technological artefacts, claiming that the mere presence of the latter influences human agents' actions (Jasanoff 2014) by linking the social and the material (Lievrouw 2014). Others focus on the affordances granted by technological artefacts and consider the invitations they provide for human agents to act (Hutchby 2001). These accounts are valuable for the understanding of algorithms and, in particular, algorithmic decision-making systems. In that regard, the design of algorithmic systems should be considered as an interface between human agents and technological systems. For instance, Allhutter et al. (2020) show in their report about the Austrian employment service (which has commissioned an ADS system for job placement) that employees need more training to react appropriately to the system and object to decisions they disagree with. Furthermore, the algorithm challenges employees' objections by demanding an explanation, while no explanation is required to accept the algorithm's decision suggestion; the act of objecting to the ADS system's recommendations is thus impeded due to the increased time requirement, and objecting reduces the amount of time available to interact with each job seeker. Therefore, the intersubjective experience – previously crucial for job placement – is critically reduced. This example shows the co-production of social realities by human agents and technological artefacts. Various contextual implications need to be considered, such as the social contexts of the human agent and the technological artefact, the affordances the artefact grants, the influence of its presence and the consequences it can entail (such as various biases). Elaborating in detail on all of these aspects, both in general and with specific regard to VERA, would go beyond the scope of this paper. However, we developed our research design and the implementation of the system against the backdrop of these considerations.

Another prominent issue outside of the scope of this paper concerns data protection and the legal framework in light of the General Data Protection Regulation (GDPR) as well as how organisations can ensure data protection while using ADS systems. The data used by such systems and the further use of information provided by users might be problematic with respect to privacy and consent. Many ADS systems rely on third-party data sources (such as social media data, data from people's smartphones or other traces data subjects leave on the internet) to increase the accuracy of the systems' recommendations (Castellucia and Métayer 2019; Lohokare, Dani, and Sontakke 2017; Wei et al. 2016) in addition to data provided by the individuals for the system with explicit consent (although even such consent can be questionable, e. g. if obtained using so-called "dark patterns" of UX design). In that regard, the GDPR provides a legal framework applicable to regulating the use of ADS systems, e. g. informed active consent (Art. 6–7), the right to be forgotten and have data deleted (Art. 17) and the right to have incorrect data rectified (Art. 16). Furthermore, the GDPR requires that data subjects are informed about ADS systems' use of their data (Art. 13) and regulates the automated processing of user data, in particular giving them the right not to be subject to purely automated decision-making (Art. 22). The regulation moreover prescribes the possibility of a human reassessment of an ADS system's decisions (Dreyer and Schulz 2019).

However, several questions remain open concerning the legislation, such as the protection of individuals in light of data controllers' opportunities to make further use of the data, for instance, to make a profit by selling them. Moreover, transparency of such complex systems both for individual users and the general public is an unresolved issue (Castets-Renard 2019). Krafft, Zweig, and König (forthcoming) have argued that legislation specific to ADM systems should be passed (supplementing the existing general data protection law) also addressing the vast differences between the various existing and possible applications of automated systems, which are currently insufficiently accounted for. Further development of existing law has proven to be particularly necessary in light of the COVID-19 pandemic, which has markedly increased the use of ADS and its dissemination across Europe (Chiusi 2020). The European Commission's "European Approach to Artificial Intelligence"[5] as well as the recently unveiled proposal for the regulation of AI[6] seem to be promising advances in this direction.

## 7. Conclusion

The introduction of ADS systems can have a substantial impact on the effectiveness of accountability and the distribution of responsibility within an organisation. Generally, using ADS systems poses a number of challenges for the decision-making process of human operators that need to be considered, such as automation bias or complacency.

We identify three types of fundamental problems that can result from changes through the introduction of ADS systems. First, after a change in tasks, employees might lack information and training to fulfil their responsibilities in meaningful ways. Second, the decision-making employee might not have the access or authority

---

[5] https://digital-strategy.ec.europa.eu/en/policies/european-approach-artificial-intelligence (accessed 15 June 2021)

[6] https://ec.europa.eu/info/strategy/priorities-2019-2024/europe-fit-digital-age/excellence-trust-artificial-intelligence (accessed 15 June 2021)

required to make necessary changes in an ADS system within their responsibility when encountering problems with or mistakes in the system. Third, through the reassignment of existing and introduction of new roles (such as development of the ADS system, system security, data protection and so on), some responsibilities might not be sufficiently clearly assigned or even not assigned at all, creating what we call a "responsibility vacuum".

For such changes to be addressed appropriately and to prepare for these challenges in advance, we suggest a comprehensive mapping of tasks and responsibilities. We have developed a set of guidelines (including guiding questions) for organisations to support such an endeavour. In addition, we have devised an interactive digital tool to reveal relations between (1) tasks and duties, (2) responsibilities and (3) access and authority. This tool can help identify potential problems in a specific organisational context and offer suggestions on which areas of concern to investigate, supporting the responsible and transparent use of ADS systems in organisational contexts.

## Acknowledgements

The first and third author were supported by the Digitalisation Fund of the Vienna Chamber of Labour through the project B-01 "Algorithms, Law and Society: Decision Makers Between Algorithmic Guidance and Personal Responsibility". We are grateful to Sabrina Burtscher, Maximilian Heimstädt, Daniel Susser, Elisa Wiedemann and the anonymous referees for suggesting numerous improvements to both the content and the presentation of this paper.

Institute of Public Law and Political Science, University of Graz, Universitätsstrasse 15/C3, 8010 Graz, Austria and Vienna Centre for Societal Security, Paulanergasse 4/8, 1040 Wien, Austria

*Email address*: angelika.adensamer@uni-graz.at

Weizenbaum Institute for the Networked Society – The German Internet Institute, Hardenbergstrasse 32, 10623 Berlin, Germany

*Email address*: rita.gsenger@rewi.hu-berlin.de

Institute of IT Security Research, St. Pölten University of Applied Sciences, Matthias-Corvinus-Strasse 15, 3100 St. Pölten, Austria

*Email address*: mail@l17r.eu

*URL*: https://l17r.eu